\documentclass{article}

\usepackage{PRIMEarxiv}

\usepackage[utf8]{inputenc} 
\usepackage[T1]{fontenc}    
\usepackage{hyperref}       
\usepackage{url}            
\usepackage{booktabs}       
\usepackage{amsfonts}       
\usepackage{nicefrac}       
\usepackage{microtype}      
\usepackage{lipsum}
\usepackage{fancyhdr}       
\usepackage{graphicx}       
\graphicspath{{media/}}     

\title{Examining the Potential for Conversational Exploratory Search using a Smart Speaker Digital Assistant
\thanks{\textit{\underline{Citation}}: 
\textbf{Kaushik, A. and J. F. Jones, G. (2023). Examining the Potential for Conversational Exploratory Search Using a Smart Speaker Digital Assistant. In Proceedings of the 18th International Joint Conference on Computer Vision, Imaging and Computer Graphics Theory and Applications - HUCAPP, ISBN 978-989-758-634-7; ISSN 2184-4321, SciTePress, pages 305-317. DOI: 10.5220/0011798700003417.}} 
}

\author{
  Abhishek Kaushik \thanks{Now at Dundalk Institute of Technology, Dundalk, Ireland.} \\
  ADAPT Centre, School of Computing \\
  Dublin City University \\
  Dublin 9, Ireland\\
  \texttt{abhishek.kaushik2@mail.dcu.ie} \\
   \And
  Gareth J. F. Jones \\
  ADAPT Centre, School of Computing \\
   Dublin City University \\
  Dublin 9, Ireland \\
  \texttt{Gareth.Jones@dcu.ie} \\
}

\begin{document}
\maketitle

\begin{abstract}
Online Digital Assistants, such as Amazon Alexa, Google Assistant, Apple Siri are very popular and provide a range or services to their users, a key function is their ability to satisfy user information needs from the sources available to them. Users may often regard these applications as providing search services similar to Google type search engines. However, while it is clear that they are in general 
able to answer factoid questions effectively, it is much less obvious how well they support less specific or exploratory type search tasks. We describe an investigation examining the behaviour of the standard Amazon Alexa for exploratory search tasks. The results of our 
study show that it not effective 
in addressing 
these types of information needs. We propose extensions to Alexa designed to overcome 
these shortcomings. Our Custom Alexa application extends Alexa's conversational functionality for exploratory search. A user study shows that our extended Alexa application both enables users to more successfully complete exploratory search tasks and is well accepted by our test users.
\end{abstract}

\keywords{Conversational search \and Exploratory Search \and Dialogue System \and Alexa.}

\section{\uppercase{Introduction}}
\label{sec:introduction}

There is currently much interest in conversational digital assistants embedded in smart speaker and mobile platforms, e.g. Amazon Alexa, Google Assistant, Apple Siri. These applications offer users a range of services including simple command and control of networked smart home appliances, accessed through conversational engagement. A widely promoted function is their ability to satisfy user information needs from the sources available to them. Digital assistants are often demonstrated using requests such as fetching recipes or latest weather forecasts. While it is clear that these application 
are often able to address such requests, which are generally satisfied by single items or factoids, it is much less clear how well current applications support more exploratory information needs, and what additional functionality might be required to address any identified shortcomings.

While conventional information retrieval (IR) systems, such as web search engines, rely on the searcher's ability to browse retrieved content in an efficient manner, smart speaker systems are largely driven by spoken interaction, sometimes involving multi-modal output. User access to returned information in spoken form has a much lower bandwidth than visual review of textual. This suggests that digital assistants must select information to be returned in spoken form in search applications with higher precision than is the case for conventional IR systems.
One way to limit delivery of extraneous information is to partition the search process into smaller incremental tasks where the searcher engages with the digital assistant using a conversational search process \cite{radlinski2017theoretical}. 

Interest in conversational search has developed rapidly the IR research community in recent years, and while it has been the focus of multiple workshops exploring its principles and challenges, there is very limited work exploring user interaction with working systems. Good examples of existing work such as \cite{trippas2017crowdsourcing,Trippas2018,trippas2017people} are limited to the use of Wizard type conversational agents in limited search tasks.
\begin{figure*}[!ht] 
    \centering
    \includegraphics[scale=0.5]{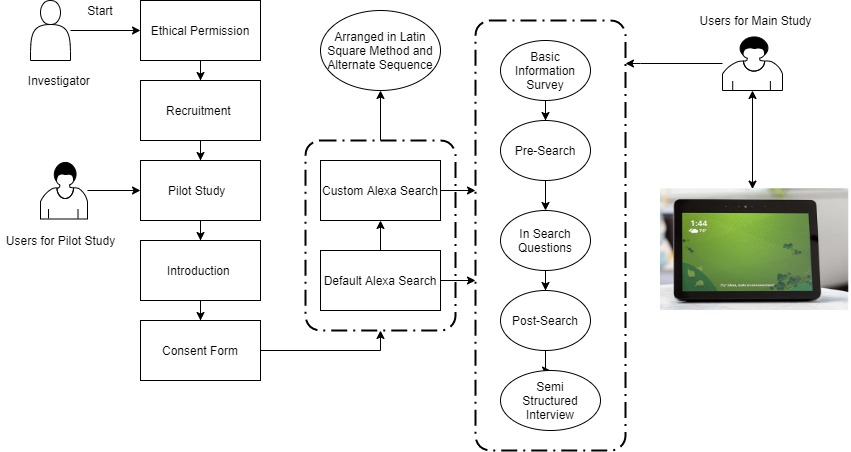}
    \vspace{-.5ex}
    \caption{Procedure of the Alexa information seeking study}
    \label{fig:myProcedure}
\end{figure*}
While in some contexts spoken only engagement is possible, the operational platforms of many digital assistants enable some form of multi-modal interaction. For example, smartphones, tablets and dedicated platforms such as the Amazon Echo Show \footnote{https://www.amazon.co.uk/amazon-echo-show-5-compact-smart-display-with-alexa/dp/B07KD7TJD6} . 

In this paper we report an experimental study of search using a state-of-the-art digital assistant to examine the ability of current assistants to satisfy exploratory information needs. For our study we adopt the Amazon Alexa operating on an Echo Show platform embedded with a display screen. The Echo Show enables conversational interaction with the Alexa assistant and incorporates a tablet sized screen to enable multi-modal engagement.
The results of our study demonstrate that the existing Alexa system is very limited in terms of its support for search tasks of this type, leading to frustration and user dissatisfaction.  We then describe experiments using a prototype new Alexa customed skill which seeks to overcome these shortcomings. Before describing these studies, the next section reviews relevant existing work in conversational search.
\begin{figure}[!ht]
    \centering
    \includegraphics[scale=0.40] {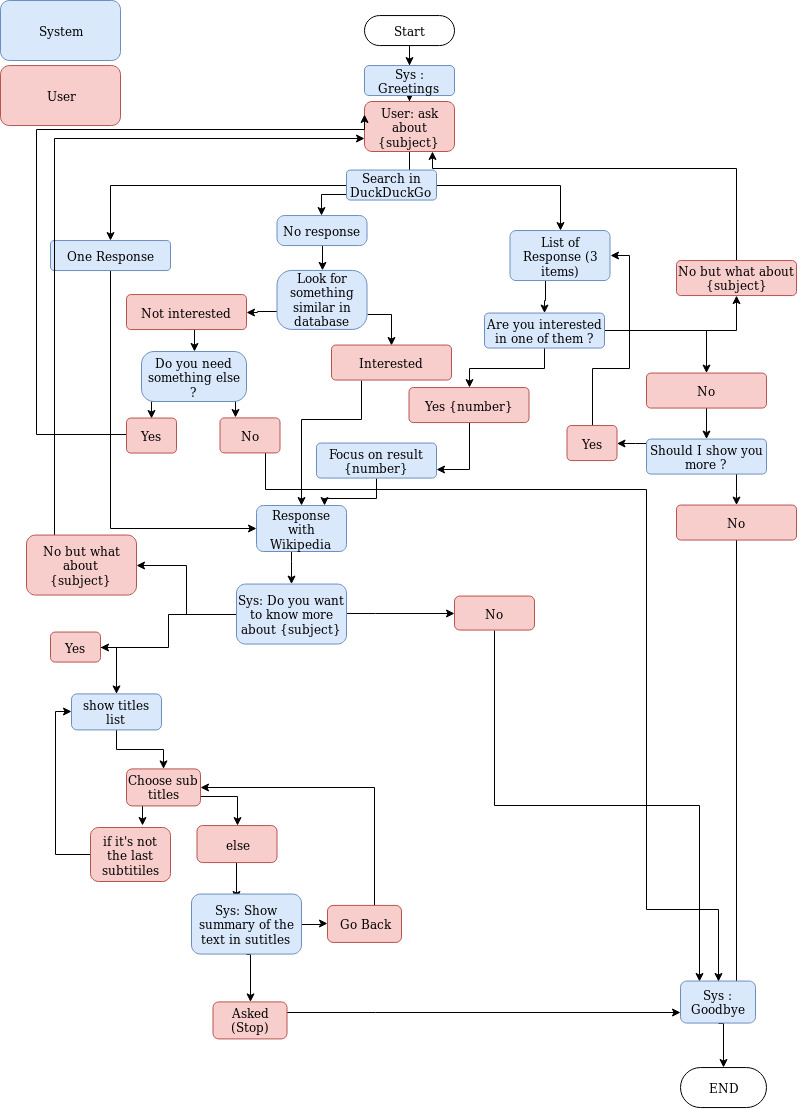}
    \caption{Flow Chart for Custom Alexa Dialogues strategy}
    \label{fig:my_Dialogues}
\end{figure}

\begin{figure}[!ht] 
    \centering
\includegraphics [scale=0.4]{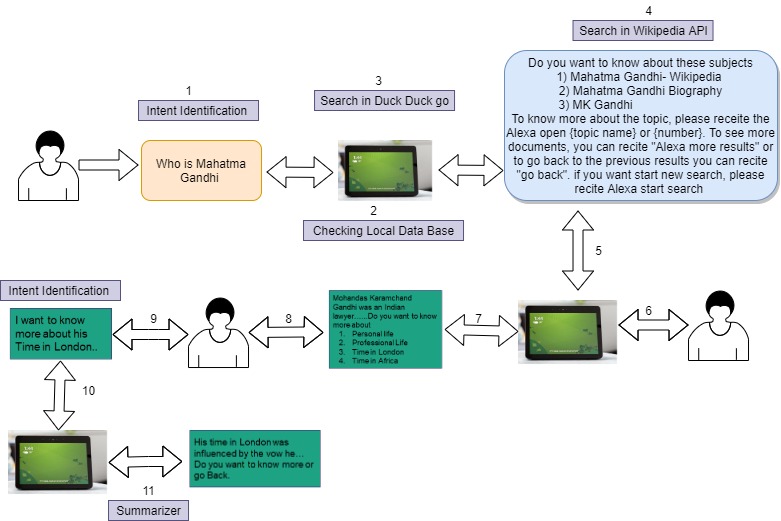}
    \caption{Alexa Custom Search}
    \label{fig:Alexa Custom Search}
\end{figure} 
\section{\uppercase{Background}}
In this section, we review literature relevant to the development of conversational search applications. We begin with a brief review of conversational systems, then consider existing work on conversational search, and look at dialogue strategies

\subsection{Conversational Systems}
Conversational engagement between human and computer applications is currently a very active area of investigation \cite{radlinski2017theoretical}. Here we focus on the area of digital conversational assistants. A study on conversational agents (Alexa vs. Siri vs. Cortona vs. Google Assistant) to investigate these usability for different services such as access to music services, agenda, news, weather, To-Do lists and maps or directions and others is reported in \cite{Lopez}. This showed that even though there are many services already available, there is much scope to improve the usability of these systems.

A review of multiple studies related to conversational agents examining their usability and the capabilities of conversational agents is described in \cite{Alexacapabilites}. A study investigating user interactions with Amazon Alexa focusing on the types of tasks requested and the variables that affect user behaviours can be found in \cite{ExploringmeAlexa}. The results indicate that across all age groups, Alexa was primarily used for checking weather, playing music, and controlling other devices. Users reported being satisfied with Alexa even when it did not produce the information sought, suggesting that the interaction experience is more important to the users than the interaction output. 

\subsection{Conversational Search}
While users of search tools have become accustomed to standard ``single shot'' interfaces, of the form seen in current web search engines, and have learned to use them to good effect, research Interest in conversational interactions in search has expanded considerably  \cite{radlinski2017theoretical}. 

Multiple studies have investigated the potential of conversational search.
However, these 
have generally involved the use of a human serving on the agent manages to the search dialogue wizard \cite{avula2018searchbots,avula2019embedding,avula2020wizard}.
While the results of these studies have been interesting and insightful, they have an important limitation in that the agent has full human intelligence. Thus, they do not reveal the potential for artificial agents to support search in terms of effectiveness and user acceptance. Studies have also been conducted to investigate the user search behaviour in speech settings where the searcher interacts with the agent (the human ``wizard'') via speech. These have the limitation of assuming both human intelligence and error free speech recognition, which will generally not be the case in a real system \cite{trippas2017crowdsourcing,trippas2017people,Trippas2018}. 


\subsection{Dialogue Strategy}
A number of frameworks have been proposed for human-machine dialogue systems. There are two important points to be considered in developing dialogue strategies. The primary one is to understand the current intention of the dialogue, and the second is to maintain the ongoing interactions between the user and the system.  The use of dialogue models is currently very limited in IR. 
In the study\cite{sitter1992modeling}, authors model information seeking dialogue as a directive, commissive and assertive type of dialogue act (e.g. asking, rejecting, offering and answering). This research was extended by Belkin et. al. \cite{belkin1995cases} to design an interactive IR system based on scripts and cases, called MERIT (Multi-Media Information System). This considers different information seeking strategies scripts and cases at once to make the system interactive. In the study \cite{leech2003generic}, the authors conducted a study in which they annotated the task-oriented dialogues in speech and introduced a dialogue taxonomy. The investigation was extended by \cite{loisel2009modeling} which proposed a model on the basis of the taxonomy and analyzing a medical text corpus.




\section{\uppercase{SEARCH USING AMAZON ALEXA}}
In this section we begin by outlining the features of the Amazon Alexa application, and then introduce the search tasks used for our investigation of its use for exploratory search.

\subsection{Amazon Alexa}

The Amazon Alexa digital assistant provides a wide range of information seeking services and control of applications to users. Alexa can operate on a range of dedicated hardware platforms including Amazon Echo, Amazon Echo Show, Amazon Dot, and related hardware, as well as an application running on more general platforms. Alexa performs voice-operated functions while communicating through a local WiFi Internet connection or other wireless connection with Amazon’s AWS cloud servers, or other networked devices, to carry out these functions \cite{janarthanam2017hands,AlexaDeveloper}.

The workflow of a standard user engagement with Alexa is divided into four steps: receiving a spoken instruction or request, interaction mode (responsible for speech recognition and intent identification), skill application logic (action after triggering the intent), and response. Where an \emph{intent} is defined by Amazon as actions that fulfill spoken requests from the user, and a \emph{skill} is an application which enables Alexa to perform an operation. A key feature of Alexa as a research tool is that new skills can be created to enable Alexa to perform new or extended operations \cite{biehl2019making}; it is for this reason that we choose to use Alexa for our investigations. Specifically, we base our study on the use of an Amazon Echo Show which combines spoken interaction with the availability of a high quality 7-inch touchscreen display which can be used within the applications to enable mulit-modal interaction. Technical details of the Echo Show can be found in 
\cite{akon2018echo}.


\subsection{Exploratory Search 
with Alexa}

\begin{figure}[!ht] 
\begin{flushleft}
{\tt Martin Luther King Jr. was an American Baptist minister and activist who became the most visible spokesperson and leader in the civil rights movement. You have to find Information about the personality using Alexa skills (as per the search setting) and based on your information gain, you have to write a short summary (in the questionnaire) about the person mentioned above and fill the questionnaire accordingly.}
\end{flushleft}
\caption{Example backstory selected for Alexa task}
\label{Alexa backstory}
\end{figure}

In this study we first examine the ability of the standard Alexa assistant to support exploratory 
information seeking using its default conversational interaction features. As a source of information needs for our study, we provide participants with backstories requiring information about an individual which we anticipate users to able to address using a single Wikipedia autobiography page. An example of backstory expressing an information need of this sort is shown in Figure \ref{Alexa backstory}).  We developed twelve backstories for which full review of the corresponding Wikipedia autobiography page is a cognitively complex task, such a task would be classified as class "Analyze" within the Taxonomy of Learning \cite{krathwohl2002revision}.

\subsection{Experimental Procedure}

Participants in our study had to complete a structured search session, as shown in Figure \ref{fig:myProcedure}. They were given printed details of the instructions for their search session, and were provided with an opportunity to familiarize themselves with using Alexa for 5-10 minutes before starting the main study. Each participant had to complete one search task using an assigned backstory. They were given the printed backstory to study before they began from the search, this was then withdrawn to prevent them simply copying the details of the backstory as the basis of their query. The search session included completing questionnaires before and after carrying out the search task. The questionnaire included asking about the 
participant's expectations and experience of the search process, and required them to write a short summary of their knowledge of the topic of backstory before and after carrying out the search process. After completing the search, they were also required to attend a semi-structured interview to gather details of their experience.

While participants carried 
the search tasks using the Echo Show, they completed the questionnaires online using a standard computer. All search activities were video recorded for post-collection review of the user activities. Approval was obtained from our university research ethics committee prior to beginning the study. Search tasks were evaluated by analyzing the self-reported questionnaire interviews and the recorded videos.  All 
details from the interviews and video recordings were assigned to response categories by independent analysts. This was done using an annotation schema relevant to our research aims designed after investigating the data; the complete response dataset was then coded using these data-derived codes \cite{braun2013successful} . The inter-rater reliability between the annotators was very high (K = 0.85 and 0.82). Search and interaction behaviour was analyzed in terms of queries used and time spent on the search task.


\subsubsection{Pilot Studies}


\begin{table}[!ht] 
    \centering
    \begin{tabular}{|c|c|c|c|}\hline
      Age & No. Male  & No. Female & Ratio  \\  
        & (M)  & (F) & (M/F) \\ \hline 
     18-25    & 14 & 3 & 14:3 \\
      26-35   & 8 & 6  & 4:3 \\
       36-45  & 0 & 0 &  NA \\ 
       Total & 22 & 9 & 22:9 \\ \hline 
    \end{tabular}
    \caption{Details of age distribution throughout this investigation}
    \label{tab:my_label_Age}
\end{table}

Prior to the main study, a pilot study was carried out by two undergraduate students in Computer Science using two additional backstory search tasks. Feedback from the pilot study was used to refine the specification of the questionnaire, and to design the classification categories for the user responses. Results from the pilot study are not included in our analysis in this project. 
Each of the pilot search tasks took around 25 minutes to complete. 

\subsubsection{Main Study Participants}

Each participant in the main study was assigned one of the selected backstories from 12 with the expectation that their session would last around 30-40 minutes. The sequence of the tasks was 
arranged using a 
Latin square method to avoid sequence and learning effects. 

In total 33 subjects participated in the experiments of which results for 2 subjects were not included for analysis due discrepancies in their data. The complete demographic details found in the Table \ref{tab:my_label_Age}. 


\section{BEHAVIOUR OF ALEXA FOR NON-FACTOID AND EXPLORATORY SEARCH}
This study examined user expectations of the Alexa assistant to support exploratory search tasks and their experiences when using 
Alexa to address this type of information need.

The following research questions were investigated in the study:
\begin{enumerate}
    \item What are the challenges and opportunities of exploratory conversational search using Alexa?
\item What characteristics of Alexa prevent it from functioning as an effective tool for complex information seeking?
\item What are the main expectations of users for conversational search systems?
\end{enumerate}

\subsection{RQ1: What are the challenges and opportunities of exploratory conversational search using Alexa?}


\subsubsection{Challenges}

Attempting to use the default Alexa assistant to address the exploratory information needs expressed in our backstories led to considerable user frustration with poor success in addressing the information need. From analysis of user feedback, we identified the following challenges.

\begin{enumerate}

\item Task Success: In 62\% of 
cases, either Alexa did not provide a response or gave irrelevant answers for the user query. 

\item User frustrations and feedback: On average participants took approximately 5.5 minutes with an average 14.1 interactions (turns). Survey feedback from the users clearly indicates high levels of user frustration.

\item 	Major limitations of Alexa in exploratory search: A number of limitations were identified from observation of user interactions and their feedback. We divided these limitations into four broad categories as follows: 

\begin{enumerate}

\item 	Poor knowledge representation: From the responses to queries given by Alexa, it became clear that Alexa only represents either fact-based answers or simply starts reading from the beginning of a long Wikipedia document.  This was noted by around 18\% of participants who claimed that Alexa had poor knowledge representation. 

\item 	Poor speech recognition and high error rate: While not directly related to its search capabilities, around 52\% of participants noted that they experienced frustration arising from poor speech recognition and high word error rates while interacting with Alexa. For example, one participant noted that \textit{``Alexa was not able to understand my voice and its frustrating and tiring to ask same thing again and again''}. Such errors can result in problems of participants being able to frame their desired query and recognition errors leading to Alexa making mistakes in interpreting the query correctly, leading to incorrect responses.

\item 	Difficulty in asking questions: This was the most important reason identified by participants, 75\% of whom indicated that they had difficulty in creating queries. They 
were unable to search effectively since they had no background knowledge about the subject. Alexa offered no formal support to them in forming queries, and its 
answers were too precise to enable the searcher to build their known of the subject, as they might with a standard web search engine.

\item 	Others: A number of other factors were identified, the key ones were "Interruptions" (2\%) and "Cognitive Load" (4\%).  Participants observed that they were unable to complete queries due to interruptions by Alexa. In these cases, Alexa took a partially completed query as finished, and interrupted in the middle of the process of entering the query, providing results which may confuse the searcher or force them to repeat or reformulate the query. This put cognitive load and strain on the participants.

\end{enumerate}

\end{enumerate}

\subsubsection{Opportunities}
Our investigations using the standard Alexa application highlight some critical areas which provide opportunities to improve exploratory search.

\begin{enumerate}
    
\item Background knowledge support and effective knowledge representations: The Alexa application provides fact-based answers, but does not support the user in learning and refining the search. After conducting this study, we propose that the user should be provided with 
information related to their search query based on facts which could help them 
to create
more effective queries.

\item	Priming, dialogue-driven approach and interactive search process: 
We observed that Alexa  
did not actively engage user in the search process, with high levels of user frustration.
To reduce the frustration and enhance search effectiveness, we propose to introduce a dialogue driven approach to the search process.

\end{enumerate}

\subsection{RQ2: What characteristics of Alexa prevent it from functioning as an effective tool for complex information seeking?}







Our investigation found that Alexa did not support information seeking more complex than simple lookup activities. This was reported by more than 45\% of participants. 
Its 
conversational agent has not been designed to support typical information seeking strategies to help the user. Two important factors in 
information seeking strategies are:
exploration and learning, which can be further subdivided into acquiring knowledge, interaction with information sources, and engagement with information sources, comparing, reasoning, analysing evaluation, discovery, planning and forecasting. 
The three major reasons for poor information seeking in standard Alexa are reported to be: lack of background knowledge due, to which the user was unable to create the right query, Alexa not being able to correctly recognise the user's query, and poor representation of knowledge by Alexa.


\subsection{RQ3: What are the main expectations of conversational search systems?}

From our study we found five major expectations of our participants for search in conversational systems. 

\subsubsection{Exploratory}

A conversational search system should provide a broad information space to the user give them the opportunity to explore a space of relevant information and to narrow the exploration to focus on addressing their information need.

\subsubsection{Content selection}


Our investigation showed that the important variables with respect to the user experience in search 
are as follows: the average number of interactions, the number of successful interactions, the number of unsuccessful interactions, the average time to complete a search task,
and the quality of the presented text. Based on our results, we can conclude that the average interaction failure rate is around 62\%, which is very high. For the total of 438 interactions the 
average total time of interactions by a user is 5.6 minutes for the default Alexa system.
We can see from the above figures that this engagement is very inefficient 
leading to demotivation and frustration for the user.

\subsubsection{Content interactions}

Searcher interactions include use of multiple Alexa skills including navigation skills, presentation skills and speech skills of conversational agents.

\begin{enumerate}

\item 	Navigation skills: A conversational search agent should support the user in navigation through the information space or the documents.

\item 	Presentation skills: The user expects presentation in different modes. More than 85\% of users considered the combination of all three dimensions (Text, Speech and Images) are required 
to present the information in the most appropriate model during the search. 

\item Speech skills: The speech skills can be classified based on multiple parameters, including speech speed, speech recognition, interruptions, speech content and its length. 
Searchers expected speech recognition to support 
standard speaking speeds, normal day-to-day length of spoken input, 
low levels of interruption and high accuracy 
speech recognition.
   
\end{enumerate}

\subsubsection{Information representations}

In our study, we found that 18\% of the searchers reported that the information represented during the search process (in default setting) was poor. 
They also found difficulty in maintaining the contextual information flow during the task. 

We observed that the various factors can refine the information representations. These 
factors are:
length of the text  shown on the screen (optimal), query relevant information, the structure of the presentation (right combination of text, images and speech) and the flow of conversation and information.

\subsubsection{Conversational properties}


Our study also indicates properties that every conversational search system should have  \cite{ChatbotMetric} \cite{Chatbotgood}.

\begin{enumerate}

 \item 	On boarding: This is the initial interaction in which the user is introduced to the system in which it explains its competencies.

 \item 	System as teacher: The user expects a system should ease their interactions by revealing its capabilities and essentially teach the user how to use the system. In our study we observed that users who have previous experience with the conversational application interacted with it for longer (7.2 minutes) than users who using it for the first or second time (5.8 minutes). 

 \item 	Incite: We observed that most of the interactions were one-way, with the system unable to engage in useful dialogue with the user. However, ideally a conversational application should engage with back-and-forth dialogue with the user to assist them in reaching goal.

 \item 	Diverge flow and course corrections: A conversational application should be robust. As such, it should be able to handle any unexpected entries from the users and use this to guide the user towards their goal.
 
\end{enumerate}

\section{DIALOGUE STRATEGIES TO SUPPORT CONTENT ENGAGEMENT}


As a result of investigations with the standard Alexa application, we sought to develop a new dialogue strategy for the Alexa assistant with the goal of improving its ability to support exploratory search. We refer to this revised Alexa application as {\em Alexa Custom search\/}. 
We implemented this as an Alexa skill designed to enable a user to carry out exploratory interactive search with Alexa.
We deployed this as a prototype using the Amazon Echo show and investigated its effectiveness using a study following the same experimental setup as used in our exploration of standard Alexa application for exploratory search.

A total 31 search sessions were conducted using same participants as the standard Alexa study, but with a different backstory assigned to each user following a Latin square backstory assignment process to avoid biasing effects between participants and assigned backstories. The participants were again given the opportunity to familiarize themselves with the application for 5-10 minutes prior to beginning the search task, had to complete pre-search and post-search questionnaires during the search session, and also to participate in a semi-structured interview at the conclusion of the search task. 
The Custom Alexa skills were developed iteratively using a series of pilot studies with informal feedback from participants prior to the formal evaluation described below.

\subsection{Dialogue Strategy} 


The dialogue strategy was designed to enable users to search and explore long retrieved documents, and to facilitate two-way interaction between Alexa and the searcher via a dialogue. The dialogue strategy has two major components: developing the skill (training Alexa based on the new skills, e.g., search and greetings) using an Amazon Developer account, and the second part to embed the search process into the dialogues by using python and AMAZON Alexa developer interface.


\medskip

\noindent {\em Developing Alexa Skills\/}: We designed two intents for the Alexa custom skill. These actions were developed to fulfill spoken requests entered by the searcher. Each intent has at least one trigger utterance, a predefined word or phrase which the user might say to invoke the intent. The intents are \textit{Greeting} (trained to answer greetings, unexpected questions and non-relevant questions with respect to a search) and \textit{Search} (trained to identify a search query pass it to the search system and to present the response from the search system to the users). Each intent was trained using likely user utterances with a corresponding response which would be expected by the users. We trained 80 alternative utterances for both intents. The utterances were collected using a small survey among a group of undergraduate students. 



\medskip

\noindent {\em Embedding the Search Process with Dialogue\/}: Alexa skills supported us to identify and classify different intents input by the user. Once Alexa identifies the input as the search intent and passes the user input to the search process which extracts the search query from the user input. The extracted query is searched based on similarity matching in the query archive and which asks for confirmation from the user regarding the search query. Each successful user-confirmed query is saved in our query archive, which we can subsequently utilise for similarity matching with fresh queries. This is to confirm that the correct query has been identified with the goal of reducing the error rate and improving the reliability of relevant search results. This acknowledgement from the user regarding confirmation of the query triggers the search process. All the responses from the search process were embedded within dialogue based on our dialogue strategy.


\subsection{Search Process}


Figures 
\ref{fig:Alexa Custom Search} and \ref{fig:my_Dialogues} illustrates the complete search process workflow. The search process triggered by the users contains the following sub-modules:

\begin{enumerate}

\item Calling Duck Duck API: The query is passed to the Duck Duck go search Application Programming Interface (API). The titles of the top 3 documents returned 
are displayed on 
the Alexa Echo Show screen.
The searcher can then select one of these 
by saying 'Open 1' or 'Open $<$document name$>$', or they 
may request more results by rejecting the displayed items by saying 'No, show me more results'. Alternatively, the searcher may change their query and restart the search process 
by saying 'Alexa start search'.
Once the user has selected an item from the displayed results, the dialogue strategy triggers the Wikipedia API. The role of the Duck Duck Go API here is essentially to identify a focused short form query which can be used for entry to the Wikipedia API to extract wikipedia documents to provide options to the users.

\item	Calling Wikipedia API: The title of the document selected by the user is passed to the Wikipedia search API. The section and subsection headings of the highest-ranking retrieved item are then shown to the user. 
The searcher can then 
select sections and subsections of the returned documents. These selected parts of the document are then summarised using the summarisation component outlined below. The document navigation options enable the searcher to explore the individual summarised parts of the document.

\item Calling the Summarizer: A summarizer is 
used to display the important content of a document section.
The Echo Show displays the summary along with further sub-options,
as shown in Figure \ref{fig:Alexa Custom Search}
The 
summarizer splits
a whole paragraph into sentences. Each sentence is considered an individual document. The "frequency" (TF) of each word in the document is calculated along with 
the inverse document frequency (IDF) of each word in the document. 
A normalised TF-IDF score of each sentence is then by summing the TF-IDF scores of each sentence and dividing 
by the word length of the sentence.The top 50\% of scoring sentences are extracted. The Density-Based Spatial Clustering of Applications with Noise (DBSCAN) clustering algorithm divides the sentences into ``clusters.'' A cosine similarity score is calculated between the number of clusters and the section name. Based on the cosine similarity score, the top 70\% of clusters are selected 
and are presented as the summary arranged in the same order as they appear in the actual paragraph. The searcher can explore further subsections or go back to the previous view. As soon as the user choses any section or sub-section to explore, the summarizer extracts its 
summary.

\end{enumerate}


\subsection{Additional functionality}	

Our Custom Alexa skill also displayed 
images associated with the displayed subsections
provided by the Wikipedia API. Images are scored using 
cosine similarity between the labels of the images and the title of the selected Wikipedia subsection. 
The image with the highest similarity is shown with the contents of the selected subsection.

\section{INVESTIGATING THE EFFECTIVENESS OF ALEXA CUSTOM SEARCH } 


\subsection{RQ4: How well does the Custom Alexa dialogue system support exploratory Search?}
The research question is divided into three sub-questions.


\subsubsection{RQ4(1): How effectively does it communicate information to the user?}

\begin{enumerate}
    
\item Dialogue Strategy: In the semi-structured interview, carried out in this second study, searchers reacted positively to 
Custom Alexa. Around 45\% of them found it helpful, 18\% found noted that it was 
interactive. 
12\% of searchers claimed that in the custom Alexa setting, Alexa provided them suggestions. Users also found setting 2 interesting, easy to understand with comfortable speed and structure unlike what they experienced in the default setting. 
    
\item Structure of representation of information: Representation of the information was one of the key criteria. We broke down the information structure into two major components: content of the document and the representation of the information.

\begin{enumerate}
    
\item Content of document: Around 87\% of participants 
were satisfied with the document information provided by 
custom Alexa.
The custom setting was able to satisfy around 85\% of the user's information needs.

\item Structure of text:  Around 77\% of searchers were satisfied with the text structured provided by 
custom alexa, while 9.7\% users were more content with the standard Alexa text structure. We can say that, the custom setting was able to satisfy 75\% of the searchers. The reason behind choosing the Alexa custom settings were: ease of use (7.5\%), ease of information seeking (37.5\%), its interactive nature (30\%) and that the information is more relevant and informative (25\%).
\end{enumerate}

\subsubsection{RQ4(2): How to verify the user understanding, satisfaction and search success in the dialogue-based exploratory search process?}

The statistically significant indicates where P$<$0.05. This signifies the experience of the users in custom search is statistically significant.

\begin{enumerate}

\item Knowledge gain and search success: 
Searchers rated (out of five) a range of variables comparing the Default setting and Custom setting of Alexa as shown in Table \ref{Comparison between Default Setting and Custom Setting}. Searchers rated Custom Alexa skills higher for most of the variables (as shown in Table \ref{Comparison between Default Setting and Custom Setting}) in comparison to the Default search setting.
    
\item Summary comparison: To verify the expansion in knowledge, we conducted a comparison of pre-task summary and post-task summary in Default and Custom Alexa settings using a standard comparison methodology. The summary comparison is based on three standard parameters named as  D-Qual, D-Intrp and D-Crit as defined explained in Table \ref{Summary Comparison Metric} and Table \ref {facSummary comparison} \cite{wilson2013comparison}. The difference between all factors in pre-search task with post-search task is greater in custom (C) setting than the default (D) setting. This indicates where searcher wrote a better summary with more facts and analysis when using 
Custom Alexa search.

\end{enumerate}


\end{enumerate}

\begin{table}[!ht] 
    \centering
    \setlength{\tabcolsep}{3pt} 
    \begin{tabular}{c|c|c|c}
  Variables &	Default  &		Custom  &		P Value 
\\  \hline Text Quality 	 &	3.1  &		3.9	 &	0.00012 \\ \hline 
Navigation Skills	 &	2.7 &		4.1 &		0.00001 \\ \hline 
Speech Skills &		3	 &	4.2 &		0.00013 \\ \hline 
Presentation Skills &		3.3 &		3.9 &		0.00070 \\ \hline 
Better Understanding  &     	3.3	 &	3.7	 &	0.85020 \\ \hline 
Knowledge Expansion	& 3.2	 &	3.9	 &	0.00146 \\ \hline 
Cognitive Engagement &		3.4 &		3.9	 &	0.16490 \\ \hline 
Search Session Success &	3.1	 &	3.7 &		0.01354 \\ \hline 
Suggesting skills &	2.5 &		3.7 &		0.00140 \\ \hline 
Alexa stop            &     	1.8	 &	1.3 &		0.28242 \\ \hline 
Ease of Multimodal      &   	3.7	 &	4 &		0.16067 \\ \hline 
    \end{tabular}
    \caption{Comparison between Default Setting and Custom Setting with statistical Testing: Two tailed T}
    \label{Comparison between Default Setting and Custom Setting}
\end{table}

\begin{table*}[!ht]
    \centering
    \begin{tabular}{l|l}
Parameter	& Definition  \\ \hline
Dqual &	Comparison of the quality of facts in the range 0-3 where \\ &  0 is irrelevant facts and 3 specific relevant facts. \\  \hline
Dintrp &	Measures the association of facts in a summary in the range 0-2 \\ &  where 0 represents no association of the  \\ &  facts and 2  that all facts in a summary are associated with each other in a meaning. \\  \hline
Dcrit	& Examines the quality of critiques of topic written by the author in range \\ &  the 0-1 where  0 represents facts are listed  with a thought or\\ &  analysis of their value and 1 where both advantages and disadvantages of the facts are given. \\  \hline
    \end{tabular}
    \caption{Summary Comparison Metric \cite{wilson2013comparison}}
    \label{Summary Comparison Metric}
\end{table*}

\begin{table}[!ht]
    \centering
    \begin{tabular}{c|c|c|c}
    \hline
Parameters  &	Default & 	Custom  &	P $<$0.05 \\
  &	Alexa & 	Alexa &\\ \hline
Dqual	& 14	& 35	& 0.0090 \\ \hline 
Dintrp	& 17 &	37 &	0.0157 \\ \hline 
Dcrit	& 15	& 21 &	0.0001 \\ \hline 

    \end{tabular}
    \caption{Summary comparison}
    \label{facSummary comparison}
\end{table}

\subsubsection{RQ4(3): Can priming help in information seeking and reducing error in conversation?}


During our interview sessions, users were asked about their experiences using the Alexa Custom setting. They answered questions relating to two dimensions: i) reasons to prefer Alexa Custom setting and ii) what are the challenges of using the Alexa Custom setting. 


\begin{enumerate}
    
\item Reasons to prefer Alexa Custom setting: The top three reasons to choose Alexa custom setting were: Navigation and Directed Search (13\%), Relevant and More Informative (21\%), and Options and Suggestions (19\%). Overall, the users found
it informative, well directed search, and that it provided options which gave them the opportunity for exploration throughout the search process.

\item Challenges of using Alexa Custom setting: Around 40\% 
of searchers were happy and did not find any challenges in using Alexa Custom setting, in contrast to the Default setting where around 95\% people found it challenging. 
The three top challenges were too many options (10\%), slow speed (10\%) and less freedom to ask (10\%). We considered these results to be positive outcomes for Alexa Custom setting since searchers found the response of Default Alexa setting too fast (i.e. spoken responses were delivered too fast to be fully comprehended) and that it was unable to provide suggestions and options during the search process.

\end{enumerate}

\subsection{RQ5: What is the user search behaviour and experience with the conversational system in an exploratory search setting?}

Our final research question focuses on comparing user behaviour patterns during an exploratory search using 
Default Alexa and Custom Alexa 
based on user interaction and self-reported answers in the questionnaires.  

\begin{enumerate}

\item Custom Alexa: Based on our analysis of user interactions (Table \ref{Comparison between Default Setting and Custom Setting}), we observe that participants found 
Alexa custom Search
more cognitively engaging than the default application. This observation implies two conclusions, that the custom application can hold the participant’s interest in the search process, and also that participants were able to learn and understand more using the custom application. Some users reported a lower level of knowledge of the topic before commencing the search task. However, they were interested in the topic, which led them to engage with relevant content with a very high of interaction during which they explored in great depth. Other users who also began with less knowledge had notably less interest in the topic, 
showed strong engagement with a limited amount of content but 
did not explore the retrieved content so widely. Other users with little initial knowledge of the topic but 
a very high interest it, 
engaged more with diverse sources of content, but with less interaction and less detailed examination of specific areas of content. The majority of users were comfortable with the multi-modality of their engagement and were satisfied with the exploratory custom search interface.

\item Default Alexa: In this setting, some users with less background knowledge of the topic and engaged repeatedly with the limited content by repeating the same queries to enhance their understanding and search experience with topic.  Other users restricted themselves to only few queries since the poor speech recognition that they experienced led to frustration. Other users expected more options and suggestions to be given by the system as per convention of their previous experiences with information retrieval systems. Generally, cognitive engagement with the system was less in comparison to the Alexa custom application. Most of the users were not very comfortable with the multi-modality available in the default setting. In general, we found that the individual pieces of information provided in default responses were not sufficient to develop a broad knowledge of the topic, resulting in poor post-search summaries.

\end{enumerate}

\section{CONCLUSIONS AND FURTHER WORK}

We have described a study examing the 
use the standard Alexa assistant 
for exploratory search tasks. This demonstrated that while it is generally able 
to answer factoid type questions quite successfully, it is not able to support the requirements of more exploratory search tasks. Our study highlighted these shortcomings in terms of the need to examine multiple retrieved items and specifically engaging with larger items in order to locate relevant information. In response to the identified shortcoming,
we proposed and implemented a customised Alexa application specifically designed to address these.
A second study examining our Custom Alexa application showed that it is able to successfully address the identified problems and is well received in terms of usability by the participants in our experimental study.

While our study shows how existing commercial conversational digital assistant applications such as Alexa can be successfully extended to support exploratory search.
this is only an initial prototype. In order to refine the features of our prototype application we need to study the different components 
in more detail to better understand the specific requirements of searchers and how these can be supported by conversational features. Further, it would be interesting to explore the possibility of the assistant capturing knowledge to which the user is exposed while carrying out an exploratory search task and using this to directly help the user in a conversational manner as they progress through the search process.  




\section*{Acknowledgement}

This work was supported by Science Foundation Ireland as part of the ADAPT Centre (Grant 13\//RC\//2106) at Dublin City University.

\bibliographystyle{unsrt}  
\bibliography{references}

\end{document}